\documentclass[twocolumn,prx,superscriptaddress]{revtex4}
\usepackage{graphics}
\usepackage{graphicx}
\usepackage{amsfonts}
\usepackage{textcomp}
\usepackage{amssymb}
\usepackage{mathrsfs}
\usepackage{amsmath}
\usepackage{color}
\usepackage{ulem}
\usepackage{float}

\begin{document}

\title{Topological Defects, Inherent Structures, and Hyperuniformity}

\author{Duyu Chen}
\email[correspondence sent to: ]{duyu@alumni.princeton.edu}
\affiliation{Materials Research Laboratory, University of California, Santa Barbara, California 93106, United States}

\author{Yu Zheng}
\affiliation{Department of Physics, Arizona State University,
Tempe, AZ 85287}

\author{Yang Jiao}
\email[correspondence sent to: ]{yang.jiao.2@asu.edu}
\affiliation{Materials Science and Engineering, Arizona State
University, Tempe, AZ 85287} \affiliation{Department of Physics,
Arizona State University, Tempe, AZ 85287}


\begin{abstract}
Disordered hyperuniform systems are exotic states of matter that completely suppress
large-scale density fluctuations like crystals, and yet possess no Bragg peaks similar to liquids or glasses. Such systems have been discovered in a variety of equilibrium and non-equilibrium physical and 
biological systems, and are often endowed with novel physical properties. While it is well known that long-range interactions are necessary to sustain hyperuniformity in thermal equilibrium at positive temperatures, such condition is not required for the realization of disordered hyperuniformity in systems out of equilibrium. However, the mechanisms associated with the emergence of disordered hyperuniformity in nonequilibrium systems, in particular inherent structures (i.e., local potential-energy minima) are often not well understood, which we will address from a topological perspective in this work. Specifically, we consider a representative class of disordered inherent structures which are constructed by continuously introducing randomly distributed topological defects (dislocations and disclinations) often seen in colloidal systems and atomic-scale two-dimensional materials. We demonstrate that these inherent structures can be viewed as topological variants of ordered hyperuniform states (such as crystals) linked through continuous topological transformation pathways, which remarkably preserve hyperuniformity. Moreover, we develop a continuum theory to demonstrate that the large-scale density fluctuations in these inherent structures are mainly dominated by the elastic displacement fields resulted from the topological defects, which at low defect concentrations can be approximated as superposition of the displacement fields associated with each individual defect (strain source). We find that hyperuniformity is preserved as long as the displacement fields generated by each individual defect decay sufficiently fast from the source (i.e., the volume integrals of the displacements and squared displacements caused by individual defect are finite) and the displacement-displacement correlation matrix of the system is diagonalized and isotropic. Our results also highlight the importance of decoupling the positional degrees of freedom from the vibrational degrees of freedom when looking for disordered hyperuniformity, since the hyperuniformity property is often cloaked by thermal fluctuations (i.e., vibrational degrees of freedom).



\end{abstract}
\maketitle

\section{Introduction}
Disordered hyperuniform (DHU) systems are exotic
states of matter \cite{To03, To18a} that lie between a perfect crystal and liquid. These systems 
are similar to liquids or glasses in that they are statistically isotropic and possess no Bragg peaks, and hence lack any conventional long-range order, and yet they completely suppress
large-scale density fluctuations like crystals and in this sense possess a hidden long-range order \cite{To03, Za09, To18a}. 
Specifically, the static structure factor $S(k)$, which is directly proportional to the scattering intensity 
measured in scattering experiments, vanishes for DHU systems in the infinite-wavelength (or
zero-wavenumber) limit, i.e., $\lim_{k\rightarrow 0}S(k) = 0$,
where $k$ is the wavenumber \cite{To03, To18a}. Here $S(k)$ is defined as 
$S(k) \equiv 1 + \rho\Tilde{h}(k)$, where $\Tilde{h}(k)$ is the Fourier transform of the total correlation
function $h(r) = g_2(r) - 1$, $g_2(r)$ is the pair correlation function, and $\rho$ is the number density of the system. Note that this definition implies that the forward scattering contribution to the
diffraction pattern is omitted. Equivalently, the local number variance $\sigma_N^2(R)\equiv \langle N^2(R)\rangle - \langle N(R) \rangle^2$ associated with a spherical observation window of radius $R$ grows more slowly than
the window volume (i.e., a scaling of $R^d$ in $d$-dimensional
Euclidean space) for DHU systems in the large-$R$ limit \cite{To03, To18a}, where $N(R)$ is the number of particles in a 
spherical window with radius $R$ randomly placed into the system. The small-$k$ scaling
behavior of $S(k) \sim k^\alpha$ dictates the large-$R$
asymptotic behavior of $\sigma_N^2(R)$, based on which all DHU
systems can be categorized into three classes:
$\sigma_N^2(R) \sim R^{d-1}$ for $\alpha>1$ (class I); $\sigma_N^2(R)
\sim R^{d-1}\ln(R)$ for $\alpha=1$ (class II); and $\sigma_N^2(R)
\sim R^{d-\alpha}$ for $0<\alpha<1$ (class III) \cite{To18a}. It is also noteworthy that the direct correlation function $c(r)$, defined via the Ornstein–Zernike relation $h(r)=c(r)+\rho c(r)\bigotimes h(r)$ (where $\bigotimes$ denotes convolution) \cite{Or14}, becomes long-ranged in the sense that it has an unbounded volume integral. This is in diametric contrast to standard thermal critical points in which $h(r)$ is long-ranged, and hence a system at a hyperuniform state is considered an ``inverted'' critical point \cite{To03}.

DHU states have been discovered in a wide spectrum of equilibrium and non-equilibrium physical and
biological systems \cite{Ga02, Do05, Za11a, Ji11, Ch14, Za11b, To15, Uc04, Ba08, Ba09, Le83, Zh15a, Zh15b, Ku11, Hu122, Dr15, He15, Ja15, We15, To08, Fe56, Ji14, Ma15, He13, Kl19, Le19a, Le19b, Ch18b, Ru19, Sa19, Sa20, Hu21, To21}; see Ref. \cite{To18a} for a thorough overview. The exotic structural features of DHU systems appear to endow such systems with novel physical properties. For example, disordered hyperuniform dielectric networks were found to possess complete photonic band gaps comparable in size to photonic crystals, while at the same time maintaining statistical isotropy, enabling waveguide geometries not compatible with photonic crystals \cite{Fl09, Ma13}. Moreover, certain disordered hyperuniform patterns have superior color-sensing capabilities, as demonstrated by avian photoreceptors \cite{Ji14}. Recent evidences also suggest that adding disorder into crystalline low-dimensional materials in a hyperuniform manner through the introduction of topological defects may enhance electronic transport in such materials \cite{Zh20, Ch21, Zh21}, which is complementary to the conventional wisdom of the landmark ``Anderson localization'' \cite{An58} that disorder generally diminishes electronic transport.

While it is well known that effective long-ranged interactions are required to drive an equilibrium many-particle system to
a hyperuniform state, this condition is not necessary to achieve hyperuniformity in systems out of equilibrium \cite{To18a}. Among the wide spectrum of 
hyperuniform nonequilibrium systems discovered previously, many fall into the category of inherent structures, i.e., local potential-energy minima associated with certain forms of interactions \cite{To18a}. For instance, a variety of maximally-random-jammed (MRJ) hard-particle packings \cite{Za11a, Za11b, Za11c, Za11d, At13, Ch14} are demonstrated to be hyperuniform; since in athermal systems increasing the density plays the same role as decreasing temperature of a molecular liquid and MRJ packings are local density maxima, these MRJ packings are considered inherent structures. The amorphous inherent structures in the quantizer problem also possess a high degree of hyperuniformity \cite{Kl19}. Interestingly, avian photoreceptor patterns are inherent structures associated with isotropic short-range hard-core repulsions between any pair of cells and isotropic long-range soft-core repulsions between pairs of cells of the subtype, and they are shown to be multihyperuniform, i.e., the photoreceptor patterns of both the total population and the individual cell types are simultaneously hyperuniform \cite{Ji14}. Another examples are the inherent structures associated with the $k$-space overlap potentials, which are shown to be hyperuniform \cite{Ba11}. It is also noteworthy that not all inherent structures are found to be hyperuniform \cite{Ba11}. For example, the inherent structures associated with the Lennard-Jones and steeply repulsive potentials are in general not hyperuniform due to the dominance of grain boundaries and vacancy defects \cite{We85, Ba11}. 

Despite the ubiquitous nature of disordered hyperuniform inherent structures, the mechanisms associated with the emergence of disordered hyperuniformity in many such systems are still not well understood. In this work, we provide a topological perspective to shed lights on this issue. In particular, we consider a representative class of disordered inherent structures in two-dimensional Euclidean space $\mathbb{R}^2$ which can be viewed as defected states of perfect triangular lattice crystal \cite{Ha78, Ne85} obtained by continuously introducing topological defects such as bound dislocations, free dislocations, and disclinations that are the key elements in the Kosterlitz-Thouless-Halperin-Nelson-Young (KTHNY) two-stage melting theory in two dimensions \cite{Ko73, Yo79, Ne79}. These defects are also commonly seen in 2D colloidal systems \cite{Za99, Wi11} and 2D semiconductors \cite{Zh20, Ch21} and play an important role in determining the physical properties of such materials.

Using various structural descriptors, we demonstrate that these inherent structures preserve the class-I hyperuniformity of the original triangular lattice crystal. We also show that disclinations result in the strongest ``degradation'' of the translational and orientational order of the crystal, followed by free dislocations and bound dislocations at comparable defect concentrations. The bond-orientational correlations in these structures rapidly decay to their long-range values over a short length scale, regardless of the defect types and concentrations. These behaviors are in stark contrast to those observed in thermally equilibrium configurations during the 2D melting process, which shows a two-step change in their translational and bond-orientational order correlations as temperature increases, corresponding to the two Kosterlitz-Thouless (KT) type transitions (i.e., solid-hexatic, and hexatic-liquid). In addition, the structures sampled from this 2D melting process are typically nonhyperuniform. These results highlight the importance of decoupling the positional degrees of freedom from the vibrational degrees of freedom and investigate inherent structures that correspond to local energy minima of the systems (i.e., positional degrees of freedom) when looking for disordered hyperuniformity, since the hyperuniformity property is often cloaked by thermal fluctuations (i.e., vibrational degrees of freedom).



Moreover, we derive a continuum theory to explain the hyperuniformity-preserving nature of the topological transformations that link the disordered inherent structures and the original hyperuniform crystals at low defect concentrations. We demonstrate that the large-scale density fluctuations in these inherent structures are mainly dominated by the elastic displacement fields resulted from the topological defects, which at low defect concentrations can be approximated as superposition of the displacement fields associated with each individual defect (strain source). Remarkably, the class-I hyperuniformity of the original crystal is preserved as long as the displacement fields of individual defects decay sufficiently fast from the source (i.e., the volume integrals of the displacements and squared displacements caused by individual defect are finite) and the displacement-displacement correlation matrix of the system is diagonalized and isotropic. In addition, the structure factor approaches zero with a universal quadratic scaling at small wavenumbers, regardless of the types and exact concentrations of topological defects. Our numerical results and theoretical analysis uncover the mechanisms underlying the emergence of disordered hyperuniformity in a wide spectrum of disordered structures, and provide insights to the discovery, design, and generation of novel disordered hyperuniform materials.


The rest of the paper is organized as follows: in Sec II, we describe the procedures to generate disordered inherent structures via continuous topological transformations from the reference triangular lattice crystal state in $\mathbb{R}^2$. In Sec. III, we employ various statistical descriptors to characterize the large-scale structural features, in particular hyperuniformity of the resulting inherent structures. In Sec. IV, we derive continuum theory to explain the class-I hyperuniformity of the inherent structures. In Sec. V, we provide concluding remarks.

\section{Realizations of disordered inherent structures containing randomly distributed topological defects}
\subsection{Dislocations and disclinations induced by topological transformations}
\begin{figure*}[t]
\begin{center}
$\begin{array}{c}\\
\includegraphics[width=0.95\textwidth]{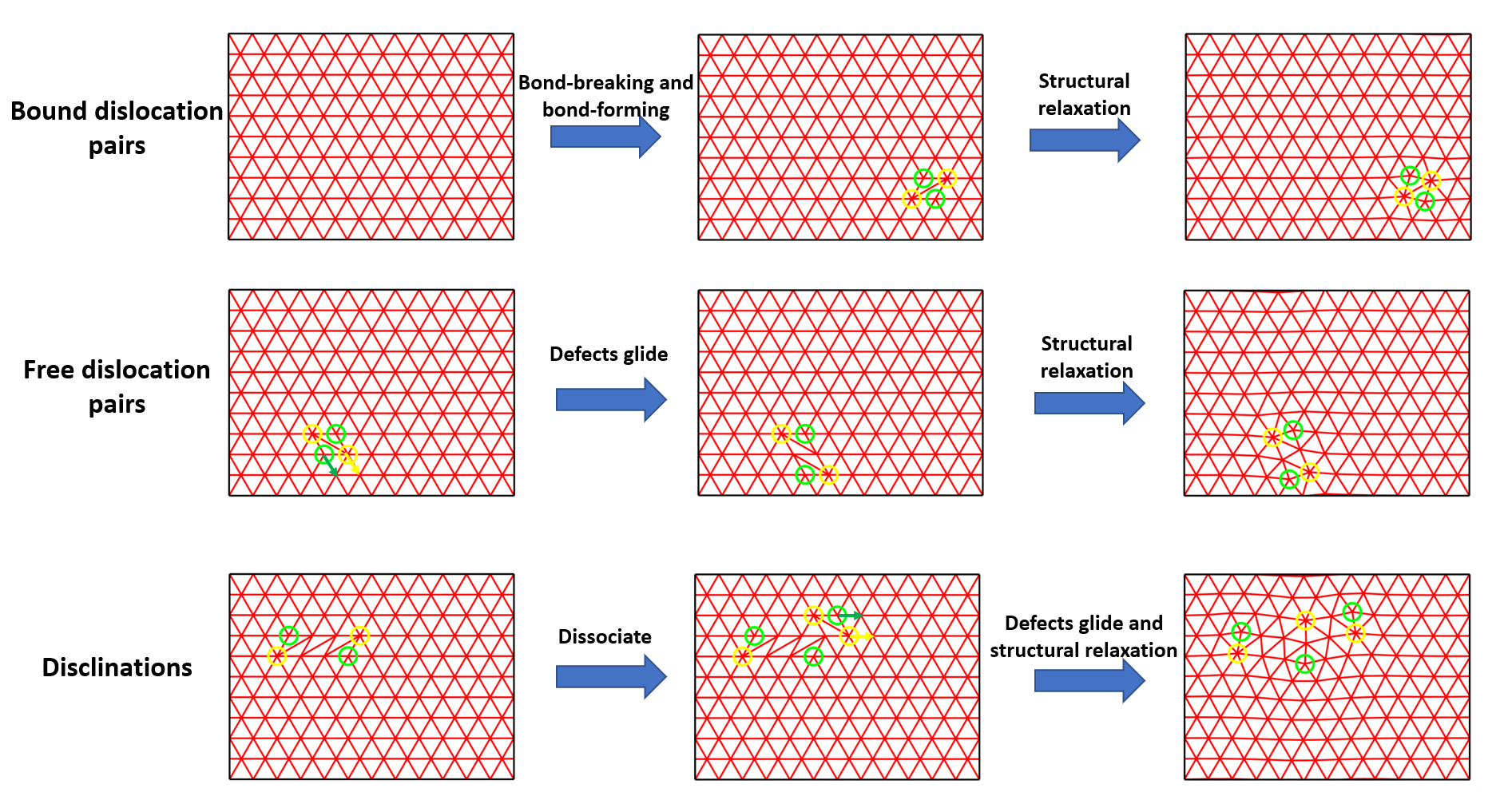} 
\end{array}$
\end{center}
\caption{(Color online) Illustration of the formation of inherent structures containing bound dislocations (top panel), free dislocations (middle panel), and disclinations with associated dislocations (bottom panel) through series of topological transformations (i.e., rearrangement of bonding network) and subsequent structural relaxation in a triangular lattice. Vertices with seven bonds are highlighted with yellow circles, and vertices with five bonds highlighted with green circles.} \label{fig_1}
\end{figure*}

To introduce bound dislocations (i.e., a pair of dislocations that are next to each other) into the triangular lattice, we first randomly pick a bond in the lattice. Note that any bond in the triangular lattice is also the short diagonal of a rhombus. Next, we break this chosen bond and connect the two vertices associated with the long diagonal of the corresponding rhombus with a new bond, resulting in a pair of dislocations next to one another. If the vertices associated with the old and new bonds all possess six bonds before the transformation, then the transformation would lead to two five-coordinated vertices and two seven-coordinated vertices; otherwise, we would obtain higher-order defected structures. Here we impose the constraint that the vertices after the transformation should each possess at leave five bonds to ensure local structural stability. The process of introducing a single pair of bound dislocations is illustrated in the top panel of Fig. \ref{fig_1}. We quantify the amount of bound dislocations by the defect concentration defined as $p\equiv N_{op}/N_b$, where $N_{op}$ is the number of successful topological transformations, and $N_b$ is the number of bonds in the triangular lattice. Note that a single topological transformation described in this paragraph would introduce a pair of dislocations, in the absence of other topological defects.

To generate free dislocations, we start from bound dislocations with the additional constraint that the initial bound dislocations should consist of two five-coordinated vertices and two seven-coordinated vertices, and randomly pick a five-coordinated vertex and a seven-coordinated vertex that are part of the bound dislocations. We then let these two defected vertices ``glide'' in the lattice by continuously breaking existing bonds and forming new bonds. Note that the direction that the defected vertices can ``glide'' is fixed once the two vertices are picked given the local bonding constraints. To form free dislocations and minimize the spatial correlations of the free dislocations, we let the defects glide for at least one step; beyond that, the defects have $1/2$ of probability to stop and $1/2$ of probability to continue gliding at each lattice site. If the gliding defects stop before hitting any ``road block'', i.e., vertices that are not six coordinated, then we count this as one successful topological transformation in the context of free dislocations. The process of introducing a single pair of free dislocations is illustrated in the middle panel of Fig. \ref{fig_1}. We also experiment with other stopping rules, and find that the details of different stopping rules do not affect the large-scale structural features of the resulting structures, which is the focus of this work. We quantify the amount of free dislocations by the defect concentration defined as $p\equiv N_{op}/N_b$, where $N_{op}$ is the number of successful topological transformations described in this paragraph, and $N_b$ is the number of bonds in the triangular lattice. Note that similar to the case of bound dislocations, a single topological transformation described in this paragraph would introduce a pair of dislocations (each consisting of a 5-coordinated vertex and a 7-coordinated vertex), in the absence of other topological defects.

To generate disclinations, we start from a free dislocation and break the bond between the seven-coordinated vertex and one of its six-coordinated neighbors and connect the long diagonal of the corresponding rhombus with a new bond. This six-coordinated neighbor should have two six-coordinated neighbors that are not neighbor of the five-coordinated vertex. This bond-breaking and bond-forming process creates an isolated 5-coordinated vertex, and another 5-coordinated vertex surrounded by two 7-coordinated vertices. We then let one of the two 7-coordinated vertices and its neighboring 5-coordinated vertex glide away in the same way as that in the case of free dislocations. If these steps can be completed, then we count this as a successful topological transformation in the context of disclinations, which would create an isolated five-fold disclination, an isolated seven-fold disclination, and two free dislocations (each consisting of a 5-coordinated vertex and a 7-coordinated vertex), in the absence of other topological defects. Such a topological transformation is illustrated in the bottom panel of Fig. \ref{fig_1}. Note that the disclinations are accompanied by free dislocations, which is consistent with previous observations \cite{Za99, Wi11} that disclinations typically arise with free dislocations. We quantify the amount of disclinations by the defect concentration defined as $q\equiv N_{op}/N_b$, where $N_{op}$ is the number of successful topological transformations described in this paragraph, and $N_b$ is the number of bonds in the triangular lattice. It is noteworthy that structures containing disclinations at $q$ should be compared to structures containing bound and free dislocations at $p=\frac{3}{2}q$ for a fair comparison at the same effective defect concentration, given the number of 5-coordinated and 7-coordinated vertices that each case generates in the absence of other topological defects.

\subsection{Inherent structures}
To obtain the inherent structures, we allow the transformed structures to undergo elastic relaxation by perturbing the positions of the vertices in a way that drive the bond lengths in the network towards values associated with the triangular lattice. In particular, this involves local minimization of the energy function $E$ defined as follows:
\begin{equation}
     E = \sum_\mathrm{bonds} k_{b} (b_i - b_0)^2,
     \label{eq_1}
\end{equation}
where $b_i$ is the bond length associated with bond $i$, and $b_0 = 1$ is the side length of a triangle in a triangular lattice. Since we are looking at local energy minima, the choice of the spring constant $k_{b}$ does not affect the obtained structure, and without loss of generality, we set $k_{b}$ to unity.

\begin{figure*}[t]
\begin{center}
$\begin{array}{c}\\
\includegraphics[width=0.95\textwidth]{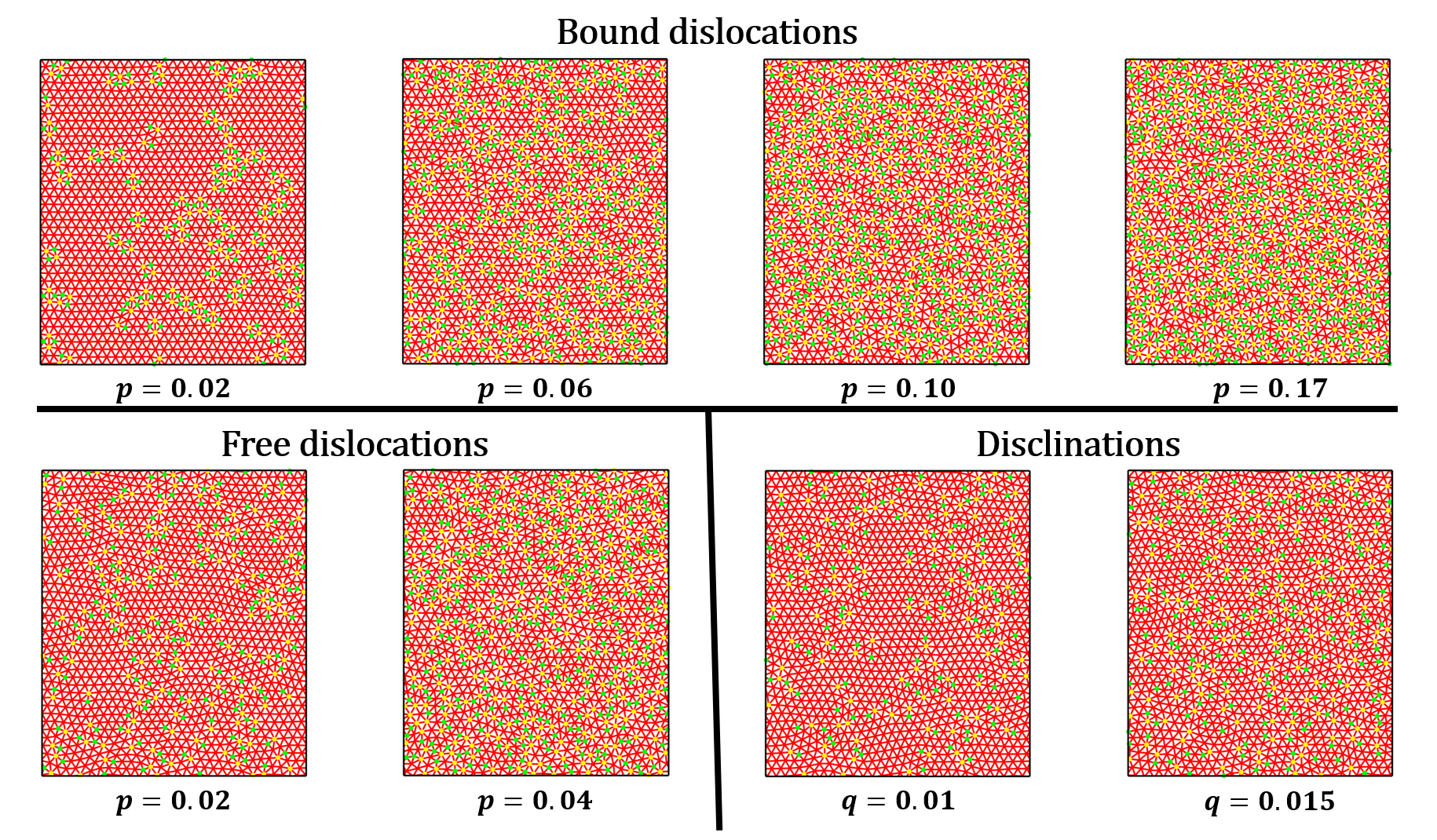} 
\end{array}$
\end{center}
\caption{(Color online) (Top section) Representative inherent structures containing primarily bound dislocations at defect concentration $p$=0.02, 0.06, 0.10, and 0.17, respectively. (Bottom left section) Representative inherent structures containing primarily free dislocations at defect concentration $p$=0.02, and 0.04, respectively. (Bottom right section) Representative inherent structures containing disclinations at defect concentration $q$=0.01, and 0.015, respectively. Vertices with more than six bonds are highlighted in yellow, and vertices with less than six bonds in green.} \label{fig_2}
\end{figure*}

We investigate inherent structures containing the aforementioned three types of topological defects for a wide range of $p$ (or $q$). In the cases of free dislocations and disclinations, we generate structures up to values close to saturation, i.e., more topological defects can no longer be inserted into the system after a sufficiently large number of attempts (e.g., $10N$ attempts, where $N$ is the number of vertices in the lattice). In the case of bound dislocations, we stop at $p=0.17$ since increasing $p$ beyond that sometimes leads to unphysical local bonding networks. In the top section of Fig. \ref{fig_2} we show representative inherent structures containing primarily bound dislocations with $N=1200$ particles, in the bottom left section of Fig. \ref{fig_2} representative inherent structures containing primarily free dislocations with $N=1200$ particles, and in the bottom right section of Fig. \ref{fig_2} representative inherent structures containing disclinations with $N=1200$ particles.

\section{Structural characterization and hyperuniformity}
To characterize the inherent structures with the aforementioned topological defects, in particular at large length scales, we generate configurations with $N=10,800$ particles at different $p$ (or $q$) and look at various statistics including pair statistics such $g_2(r)$, $S(k)$ and $\sigma_N^2(R)$ \cite{To03, To18a}, and bond-orientational statistics such as the bond-orientational order metric $Q_6$ and correlation function $C_6(r)$ that have been routinely used to study the 2D melting process \cite{Za99, Wi11, Li18}. To compute the statistics accurately, we average over 10 configurations at each $p$ (or $q$). 

Specifically, the pair correlation function $g_2(r)$ is proportional to the probability density function of finding two centers separated by distance $r$ \cite{To02a}, and in practice is computed via the relation
\begin{equation}
\label{eq_2}
 g_2(r) = \frac{\langle N(r)\rangle}{\rho 2\pi r\Delta r},
\end{equation}
where $\langle N(r)\rangle$ is the average number of particle centers that fall into the circular ring at distance $r$ from a central particle center (arbitrarily selected and averaged over all particle centers in the system), $2\pi r\Delta r$ is the area of the circular ring, and $\rho$ is the number density of the system \cite{To02a, At13}. The static structure factor $S(k)$ is the Fourier counterpart, and for computational purposes, $S({k})$ is the angular-averaged version of $S({\bf k})$, which can be obtained directly from the particle positions ${\bf r}_j$, i.e.,
\begin{equation}
\label{eq_3}
S({\bf k}) = \frac{1}{N} \left |{\sum_{j=1}^N \exp(i {\bf k} \cdot
{\bf r}_j)}\right |^2 \quad ({\bf k} \neq {\bf 0}),
\end{equation}
where $N$ is the total number of points in the system \cite{Za09, At13, Ch14}. The trivial forward scattering contribution (${\bf k} = 0$) in Eq. \ref{eq_3} is omitted, which makes Eq. \ref{eq_3} completely consistent with the aforementioned definition of $S(k)$ in the ergodic infinite-system limit \cite{To18a}. To compute $\sigma_N^2(R)$, we randomly place circular observation windows with radius $R$ in the system, and count the number of particles $N(R)$ that fall into the observation window, which is a random variable. The variance associated with $N(R)$ is denoted by $\sigma_N^2(R)\equiv\langle N(R)^2\rangle-\langle N(R)\rangle^2$, which measure density fluctuations of particles within a window of radius $R$. In this work we sample 100,000 windows at each window radius $R$ to obtain $\sigma_N^2(R)$. 

On the other hand, the order metric $Q_6$ is defined as 
\begin{equation}
\label{eq_4}
Q_6\equiv |\langle \Psi_6 \rangle|, 
\end{equation}
where 
\begin{equation}
\label{eq_5}
\Psi_6({\bf r}_i) = \frac{1}{n_i}\sum_{j=1}^{n_i}e^{6\theta_{ij}},
\end{equation}
and $\langle\cdots\rangle$ denotes ensemble average, $n_i$ is the number of neighbors of vertex $i$ located at ${\bf r}_i$, and $\theta_{ij}$ is the polar angle associated with the vector from vertex $i$ to the $j$-th bonded neighbor of vertex $i$. The bond-orientational correlation function $C_6(r)$ is defined as 
\begin{equation}
\label{eq_6}
C_6(r)\equiv\langle \Psi_6({\bf r}_i)\Psi^*_6({\bf r}_j)\rangle\mid r=|{\bf r}_i-{\bf r}_j|,
\end{equation} 
where $\Psi^*_6$ is the complex conjugate of $\Psi_6$. In practice, to calculate $C_6(r)$, for each pair of particles located at ${\bf r}_i$ and ${\bf r}_j$, we compute $\Psi_6({\bf r}_i)\Psi^*_6({\bf r}_j)$, and bin the results according to the distance $r=|{\bf r}_i-{\bf r}_j|$. We note that $Q_6 = 1$ and $C_6(r) = 1$ for a perfect triangular network; while for isotropic fluid phase, $Q_6 \approx 0$ and $C_6(r)$ decays with an exponential envelop at large $r$ \cite{Za99, Wi11}.

\subsection{Bound dislocations}
\begin{figure*}[t]
\begin{center}
$\begin{array}{c}\\
\includegraphics[width=0.95\textwidth]{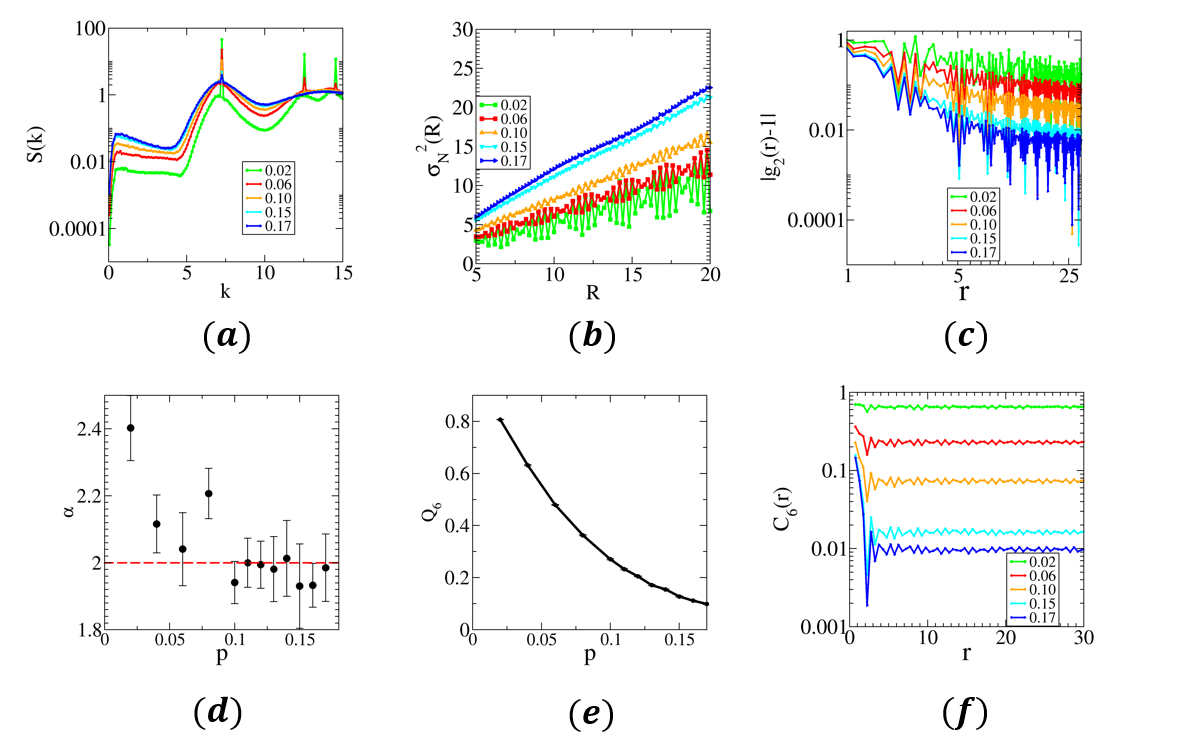} 
\end{array}$
\end{center}
\caption{(Color online) Statistics associated with inherent structures containing primarily bound dislocations at different defect concentrations $p$ with N=10,800 particles. (a) Structure factor $S(k)$. (b) Local number variance $\sigma_N^2(R)$. (c) Log-log plot of $|g_2(r)-1|$. (d) Small-wavenumber scaling exponent $\alpha$ of $S(k)$. (e) Bond-orientational order metric $Q_6$. (f) Bond-orientational order correlation function $C_6(r)$.} \label{fig_3}
\end{figure*}

We first present the results of the inherent structures with primarily bound dislocations at different $p$. In particular, as shown in Fig. \ref{fig_3}(a)-(c), the structure factor $S(k)$ decreases to essentially zero as $k$ approaches zero and local number variance $\sigma_N^2(R)$ grows roughly linearly as $R$ increases at large $R$, indicating the hyperuniformity of these inherent structures. The pair correlation function $g_2(r)$ decays to its long-range value over a short range of $r$, and the long-range value of $|g_2(r)-1|$ and the magnitudes of the Bragg peaks in $S(k)$ also decrease significantly as $p$ increases, indicating the possible loss of translational order in these systems as dislocations are introduced into the system. However, we note that the absence of Bragg peaks alone does not guarantee that the underlying structure is truly amorphous, since long-range order can be hidden at the two-point level, but still can be present at higher-point levels, as explicitly demonstrated by Klatt et al. in the context of random, uncorrelated displacements of particles on a lattice \cite{Kl20}. There are clear wiggles in $S(k)$ at large $k$ and significant oscillations in $g_2(r)$ as well, which are manifestations of the remaining short-range structures in the defected networks. We further analyze the small-wavenumber behavior of $S(k)$, and find that the exponent $\alpha$ in $S(k) \sim k^\alpha$ oscillates around 2, as shown in Fig. \ref{fig_3}(d), demonstrating that bound dislocations preserve the class-I hyperuniformity of the triangular lattice.

We also analyze the bond-orientional order of the inherent structures. The results of $Q_6$ and $C_6(r)$ are shown in Fig. \ref{fig_3}(e) and \ref{fig_3}(f), respectively. It can be clearly seen that $Q_6$ decreases rapidly as $p$ increases, indicating the loss of the global preferred orientation of the lattice. On the other hand, $C_6(r)$ decays to its long-range value rapidly over a short length scale regardless of $p$, which can be attributed to the fact that the bound dislocations are randomly introduced in the system, and the spatial correlations of defect positions are minimized. The long-range value of $C_6(r)$ also decreases as $p$ increases, indicating the loss of large-scale orientational correlation as bound dislocations are introduced.

\subsection{Free dislocations}
\begin{figure*}[t]
\begin{center}
$\begin{array}{c}\\
\includegraphics[width=0.95\textwidth]{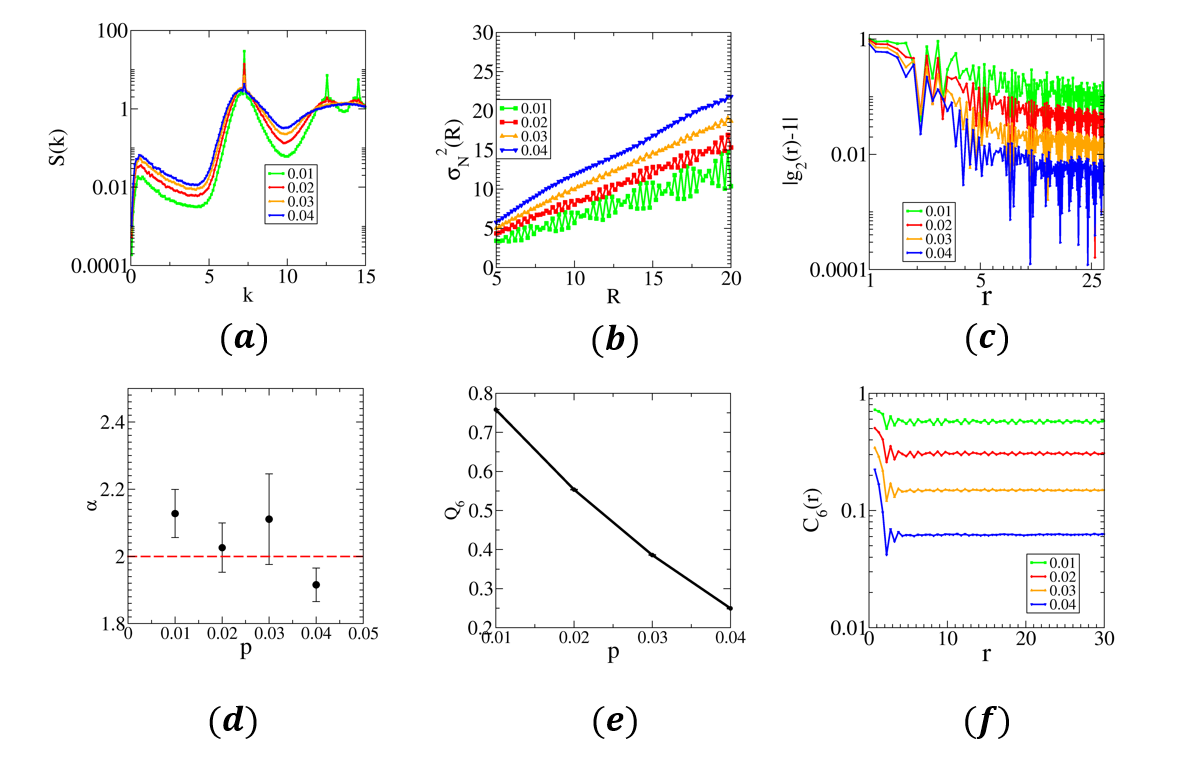} 
\end{array}$
\end{center}
\caption{(Color online) Statistics associated with inherent structures containing primarily free dislocations at different defect concentrations $p$ with N=10,800 particles. (a) Structure factor $S(k)$. (b) Local number variance $\sigma_N^2(R)$. (c) Log-log plot of $|g_2(r)-1|$. (d) Small-wavenumber scaling exponent $\alpha$ of $S(k)$. (e) Bond-orientational order metric $Q_6$. (f) Bond-orientational order correlation function $C_6(r)$.} \label{fig_4}
\end{figure*}

We employ similar procedures to investigate inherent structures containing primarily free dislocations, and the computed statistics are shown in Fig. \ref{fig_4}. Interestingly, there are many structural similarities that these inherent structures share with inherent structures containing primarily bound dislocations. For example, these inherent structures also preserve the class-I hyperuniformity of the triangular lattice, as manifested by the fact that $S(k)$ essentially decreases to zero with an approximately quadratic scaling as $k$ approaches zero and $\sigma_N^2(R)$ increases linearly as $R$ increases at large $R$. Both $g_2(r)$ and $C_6(r)$ decay to their respective long-range values over a short length scale, and $Q_6$ decreases rapidly as $p$ increases, indicating the loss of large-scale structural order as free dislocations are introduced into the system. However, we note that at the same $p$, free dislocations degrade the translational and orientational order of the triangular lattice much more than bound dislocations, as evidenced by $g_2(r)$, $Q_6$, and $C_6(r)$. This is not surprising that the impact of bound dislocations are much more localized than that of free dislocations.

It is noteworthy that in colloidal systems during 2D melting, as free dislocations begin to emerge, the systems start to enter the hexatic phase regime, and the $h(r)$ and $C_6(r)$ typically show an exponential and an algebraic decay, respectively \cite{Za99, Wi11}. These behaviors are distinctly different from those of our inherent structures containing primarily free dislocations, where $h(r)$ and $C_6(r)$ decay to their respective long-range values over a short range of $r$ and oscillate around certain constants afterwards. These differences may be attributed to the fact that in our systems, the free dislocations are introduced in a mostly uncorrelated manner, while in those colloidal systems during 2D melting, the free dislocations arise as a result of thermal excitation and possess certain degrees of spatial correlation. Our results suggest that not only the types of topological defects, but also the spatial correlation of topological defects affect the structural behaviors of the defected lattices.

\subsection{Disclinations}
\begin{figure*}[t]
\begin{center}
$\begin{array}{c}\\
\includegraphics[width=0.95\textwidth]{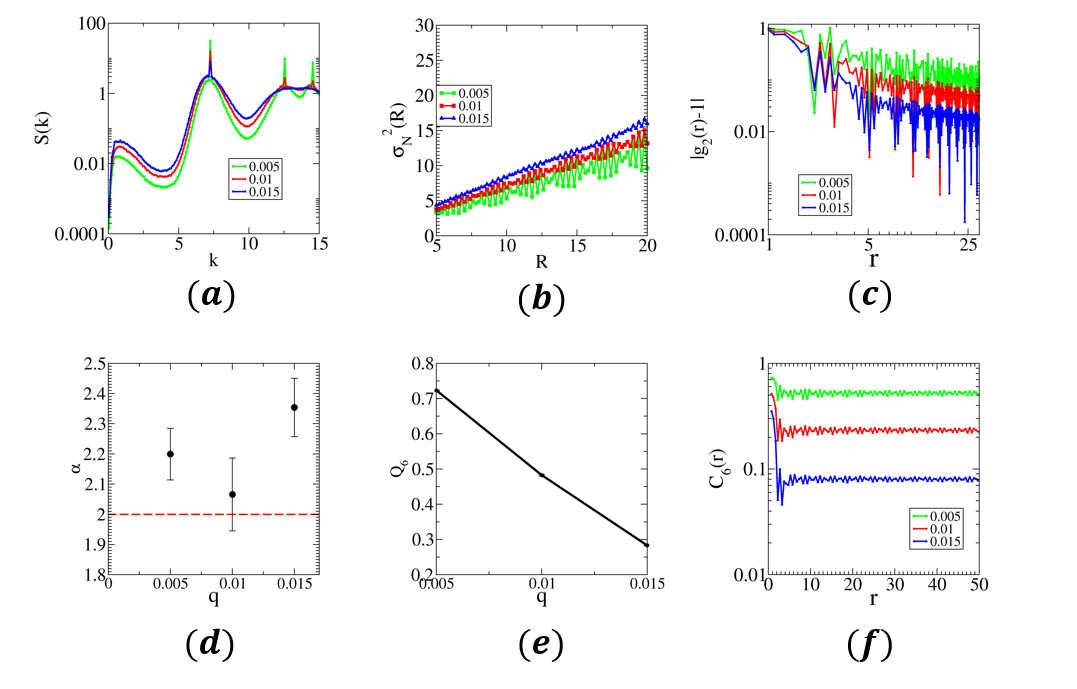} 
\end{array}$
\end{center}
\caption{(Color online) Statistics associated with inherent structures containing disclinations at different defect concentrations $p$ with N=10,800 particles. (a) Structure factor $S(k)$. (b) Local number variance $\sigma_N^2(R)$. (c) Log-log plot of $|g_2(r)-1|$. (d) Small-wavenumber scaling exponent $\alpha$ of $S(k)$. (e) Bond-orientational order metric $Q_6$. (f) Bond-orientational order correlation function $C_6(r)$.} \label{fig_5}
\end{figure*}

Next, we investigate inherent structures containing primarily isolated disclinations, and the results are shown in Fig. \ref{fig_5}. Clearly, the inherent structures preserve class-I hyperuniformity of the triangular lattice, and the translational and orientional order of the system are greatly degraded by the introduced disclinations. Remarkably, both $h(r)$ and $C_6(r)$ decay to their respective long-range values over a short range of $r$ and oscillate around certain constants afterwards. These results are surprising since previously isolated disclinations caused by thermal excitation as temperature increases in colloidal systems were known to induce large-scale structural distortions, and lead the systems to transition into isotropic liquids, which essentially lose translational and orientational order, and are generally not hyperuniform \cite{Za99, Wi11}. In particular, in those systems $h(r)$ and $C_6(r)$ both exhibit an exponential decay as $r$ increases \cite{Za99, Wi11}. These different behaviors can be attributed to the fact that in our systems the disclinations are randomly placed into the systems, which do not affect large-scale density fluctuations or orientational correlations. Nonetheless, we find that in our systems disclinations degrade the translational and orientational order much more than bound and free dislocations at comparable defect concentration, which is not surprising given that disclinations cause larger-scale structural distortions than dislocations.

\section{Continuum theory of hyperuniformity in inherent structures containing topological defects}

In this section, we devise a continuum theory to explain our observations from Sec. III of the impact of the topological defects on hyperuniformity, i.e., how the topological transformations involving dislocations and disclinations preserve the class-I hyperuniformity of the original triangular-lattice crystal? We note that the introduction of topological defects preserves the total number of particles in the system, i.e., no particles were removed or added. Therefore, the impacts on the local number density fluctuations are resulted from the perturbation of particle positions at the core of the defects and the associated elastic displacement field.  Specifically, we assume that the particle displacement (at low defect concentrations) at position $\mathbf{x}$ is the linear superposition of the displacements introduced by different topological defects at $\mathbf{r}_1$, $\cdots$, $\mathbf{r}_M$, where $M$ is the number of topological defects, i.e., 
\begin{equation}
    \label{eq_7}
    \mathbf{u}(\mathbf{x}) = \sum_{i=1}^M \mathbf{f}(\mathbf{x}-\mathbf{r}_i).
\end{equation}
Therefore, the average displacement field $\langle \mathbf{u}(\mathbf{x}) \rangle$ is given by
\begin{equation}
\label{eq_8}
\begin{split}
  \langle \mathbf{u}(\mathbf{x}) \rangle&=\int\sum_{i=1}^M \mathbf{f}(\mathbf{x}-\mathbf{r}_i)P_M(\mathbf{r}^M)d\mathbf{r}^M \\
    &=\int \mathbf{f}(\mathbf{x}-\mathbf{r}_1) \rho_{1s}(\mathbf{r}_1)d\mathbf{r}_1\\
    &=\rho_s \int \mathbf{f}(\mathbf{r})d\mathbf{r},
\end{split}
\end{equation}
where $P_M(\mathbf{r}^M)$ is the probability density function \cite{To03} associated with finding defects $1, 2, \cdots, M$ at position $\mathbf{r}_1$, $\mathbf{r}_2$, $\cdots$, $\mathbf{r}_M$, and $\rho_{ms}(\mathbf{r}^m) (m<M)$ is the reduced generic density function \cite{To03} of the defects defined as
\begin{equation}
\label{eq_9}
    \rho_{ms}(\mathbf{r}^m) = \frac{M!}{(M-m)!}\int\cdots\int P_M(\mathbf{r}^M) d\mathbf{r}^{M-m},
\end{equation}
and because of statistical homogeneity, the one-point density function $\rho_{1s}(\mathbf{r}_1)$ is equal to the average defect density $\rho_s$ in the system. Similarly, the different components of the displacement-displacement correlation $\mathbf{\Psi}_{\mu\nu}(\mathbf{r}=\mathbf{y}-\mathbf{x})\equiv \langle u_{\mu}(\mathbf{x})u_{\nu}(\mathbf{y})\rangle-\langle u_{\mu}(\mathbf{x})\rangle\langle u_{\nu}(\mathbf{y})\rangle$ are given by
\begin{equation}
\label{eq_10}
\begin{split}
  \Psi_{\mu\nu}(\mathbf{r})&=\int\sum_{i=1}^M\sum_{j\neq i}^M f_{\mu}(\mathbf{x}-\mathbf{r}_i) f_{\nu}(\mathbf{y}-\mathbf{r}_j)P_M(\mathbf{r}^M)d\mathbf{r}^M \\
    &+\int\sum_{i=1}^M f_{\mu}(\mathbf{x}-\mathbf{r}_i) f_{\nu}(\mathbf{y}-\mathbf{r}_i)P_M(\mathbf{r}^M)d\mathbf{r}^M \\
    &-\int \rho_s^2 f_{\mu}(\mathbf{x}-\mathbf{r}_1) f_{\nu}(\mathbf{y}-\mathbf{r}_2)d\mathbf{r}_1d\mathbf{r}_2 \\
    &=\int \rho_s^2 h_s(\mathbf{r}_2-\mathbf{r}_1) f_{\mu}(\mathbf{x}-\mathbf{r}_1) f_{\nu}(\mathbf{y}-\mathbf{r}_2)d\mathbf{r}_1d\mathbf{r}_2\\
   &+\int \rho_s f_{\mu}(\mathbf{x}-\mathbf{r}_1) f_{\nu}(\mathbf{y}-\mathbf{r}_1)d\mathbf{r}_1,
\end{split}
\end{equation}
where $h_s(\mathbf{r})\equiv g_{2s}(\mathbf{r})-1=[\rho_{2s}(\mathbf{r})-\rho_s^2]/\rho_s^2$ is the total correlation function of the topological defects. If the topological defects are randomly introduced into the system, then $h_s(\mathbf{r})=0$, which gives
\begin{equation}
\label{eq_11}
\begin{split}
  \Psi_{\mu\nu}(\mathbf{r})&=\int \rho_s f_{\mu}(\mathbf{x}-\mathbf{r}_1) f_{\nu}(\mathbf{y}-\mathbf{r}_1)d\mathbf{r}_1 \\
  &=\int \rho_s f_{\mu}(\mathbf{r}_1) f_{\nu}(\mathbf{r}_1+\mathbf{r})d\mathbf{r}_1 
\end{split}
\end{equation}
In the Fourier space, this corresponds to 
\begin{equation}
\label{eq_12}
    \tilde{\Psi}_{\mu\nu}(\mathbf{k}) = \rho_s \tilde{f}_{\mu}(\mathbf{k})\tilde{f}_{\nu}^*(\mathbf{k})=\rho_s \tilde{f}_{\mu}(\mathbf{k})\tilde{f}_{\nu}(-\mathbf{k}),
\end{equation}
where $\tilde{\mathbf{\Psi}}$, $\tilde{\mathbf{f}}$ are the Fourier transforms of $\mathbf{\Psi}$ and $\mathbf{f}$, respectively, and $\tilde{\mathbf{f}}^*$ is the complex conjugate of $\tilde{\mathbf{f}}$.

Next, we derive the expression for the structure factor $S(k)$ of triangular lattice affected by displacement fields $\mathbf{u}$. Previously, it was shown \cite{Ki18} in general that when the displacement field has a finite variance $\langle |\mathbf{u}|^2\rangle$ and the displacement-displacement correlation matrix is isotropic and diagonalized, i.e., $\Psi_{\mu\nu}(\mathbf{r}) = \delta_{\mu\nu} \Psi(\mathbf{r})$, the structure factor $S(\mathbf{k})$ of a displaced hyperuniform point pattern at small $|\mathbf{k}|$ is approximated by
\begin{equation}
\label{eq_13}
\begin{split}
  S(\mathbf{k})&\approx[|\mathbf{k}|^2\Psi(\mathbf{0})+(1-|\mathbf{k}|^2\Psi(\mathbf{0}))S_0(\mathbf{k})] \\
  & + \rho_0|\mathbf{k}|^2(\tilde{\Psi}(\mathbf{k})+\int d(\mathbf{r})h_0(\mathbf{r})\Psi(\mathbf{r})e^{-i\mathbf{k}\cdot\mathbf{r}})
\end{split}
\end{equation}
where $S_0(\mathbf{k})$ and $h_0(\mathbf{r})$ are the structure factor and total correlation function of the original point patterns. In the cases where the original point patterns are crystals, we can use the properties of crystals to simplify Eq. \ref{eq_13}. In particular, the structure factor $S_0(\mathbf{k})=0$ holds for $|\mathbf{k}|<K$, where $K$ is the wavenumber associated with the first Bragg peaks, and pair correlation function $g_{20}(\mathbf{r})=h_0(\mathbf{r})+1$ is simply a collection of $\delta$-functions at lattice sites and zero otherwise. By Taylor-expanding the second line of Eq. \ref{eq_13} at small $k$ and invoking the continuum approximation, we obtain the following expression:
\begin{equation}
\label{eq_14}
\begin{split}
  S(k)&\approx k^2\Psi(0)+\rho_0k^2\tilde{\Psi}(0) \\
  &= \rho_sk^2\int  f_{1}^2(\mathbf{r})d\mathbf{r}+\rho_0\rho_sk^2|\tilde{f}_1(0)|^2 \\
  &= \rho_sk^2\int  f_{2}^2(\mathbf{r})d\mathbf{r}+\rho_0\rho_sk^2|\tilde{f}_2(0)|^2
\end{split}
\end{equation}

Interestingly, Eq. \ref{eq_14} suggests that as long as the volume integrals of $\mathbf{f}(\mathbf{r})$ and $|\mathbf{f}(\mathbf{r})|^2$ are finite, the structure factor $S(k)$ of the triangular lattice affected by the displacement fields generated by the collection of randomly distributed source functions $\mathbf{f}$ scales as $S(k) \sim k^2$, which indicates that such displacement fields preserve the class-I hyperuniformity of the original crystals. 

\begin{figure}[t]
\begin{center}
$\begin{array}{c}\\
\includegraphics[width=0.45\textwidth]{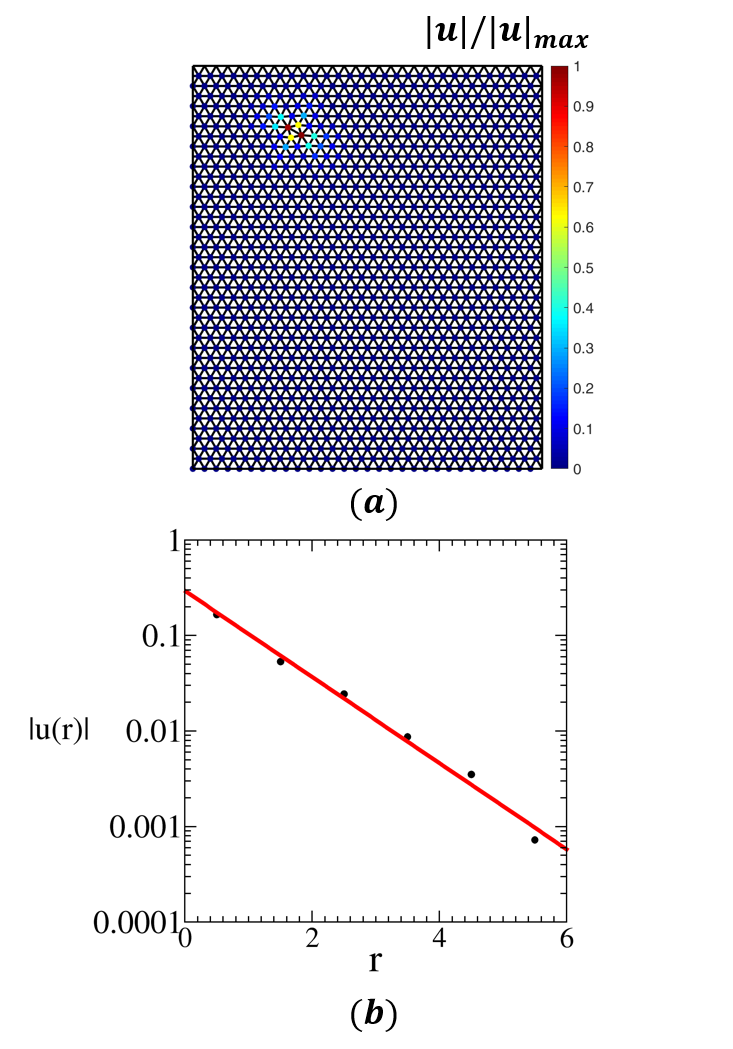} 
\end{array}$
\end{center}
\caption{(Color online) Displacement field $\mathbf{u}$ in an inherent structure containing a single pair of bound dislocations. (a) Spatial distribution of normalized $|u(r)|/|u(r)|_{max}$. (b) Decay of $|\mathbf{u}(r)|$ as a function of the distance $r$ from the core of the topological defect, i.e., the center of the old broken bond (the same as the center of the new formed bond), and the red solid line is an exponential fit (note that the vertical axis is log-scale).} \label{fig_6}
\end{figure}

Subsequently, we test our continuum theory against numerical examples of inherent structures investigated in Sec. III. We first check whether the assumptions of our theory are satisfied in these cases. In Fig. \ref{fig_6}(a) we visualize the magnitude of the displacement field $\mathbf{u}$ in an inherent structure containing a single pair of bound dislocations. Clearly, the displacement field concentrates around the center of the topological defect, i.e., the center of the old broken bond (which is the same as the center of the new formed bond). We further compute the decay of $|\mathbf{u}(r)|$ as a function of the distance $r$ from the core of the topological defect, which appears to decay exponentially as shown in Fig. \ref{fig_6}(b), although we note that $|\mathbf{u}(\mathbf{r})|$ appears to be anisotropic around the core as shown in Fig. \ref{fig_6}(a). The exponential decay suggests that the volume integrals of the source $\mathbf{f}$ and $|\mathbf{f}|^2$ should be finite. 

\begin{figure*}[t]
\begin{center}
$\begin{array}{c}\\
\includegraphics[width=0.8\textwidth]{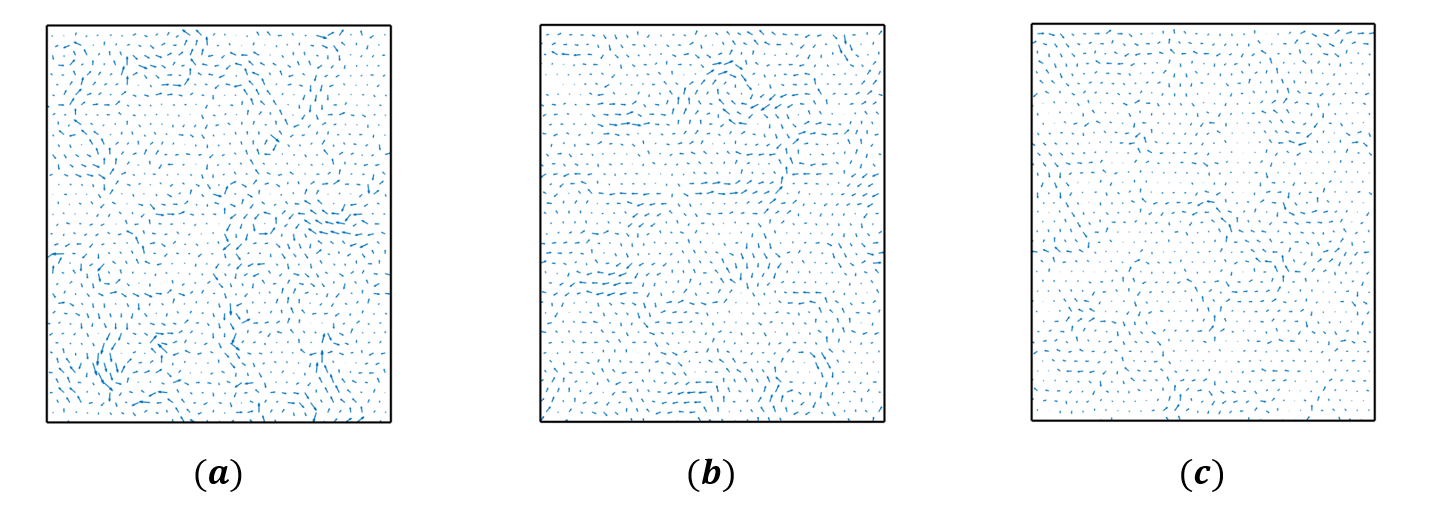} 
\end{array}$
\end{center}
\caption{(Color online) (a) Spatial distribution of the vector displacement field $\mathbf{u}(\mathbf{r})$ in an inherent structure containing primarily bound dislocations at $p=0.17$ with $N=1200$ particles. (b) Spatial distribution of the vector displacement field $\mathbf{u}(\mathbf{r})$ in an inherent structure containing primarily free dislocations at $p=0.04$ with $N=1200$ particles. (c) Spatial distribution of the vector displacement field $\mathbf{u}(\mathbf{r})$ in an inherent structure containing disclinations at $q=0.015$ with $N=1200$ particles.} \label{fig_7}
\end{figure*}

\begin{figure*}[t]
\begin{center}
$\begin{array}{c}\\
\includegraphics[width=0.9\textwidth]{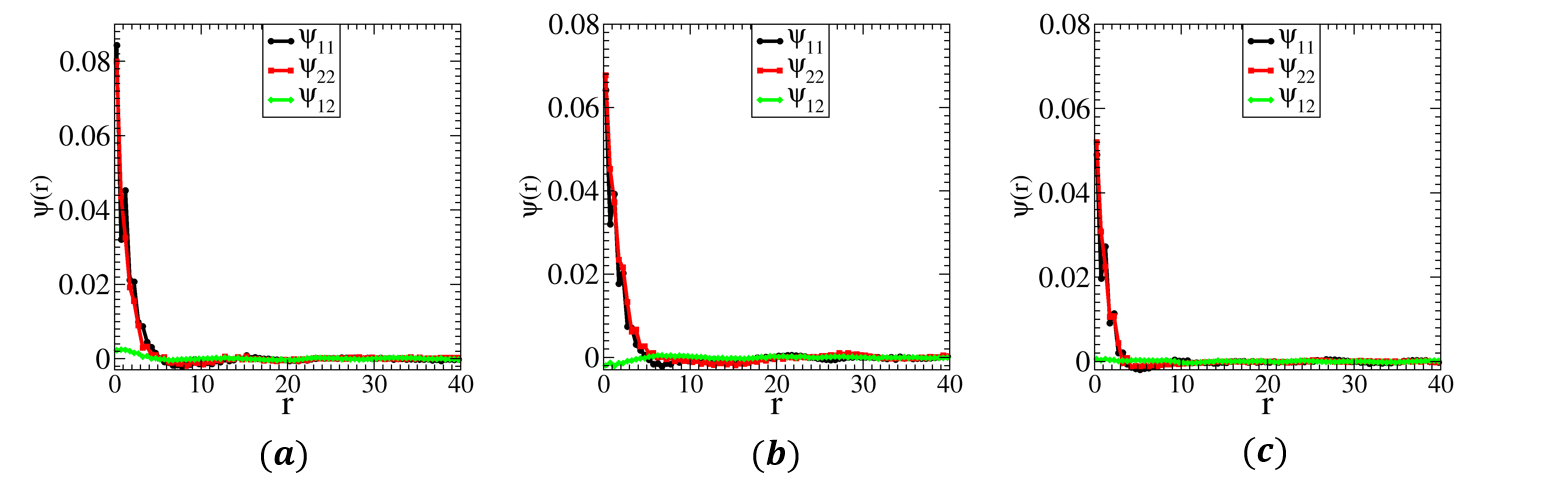} 
\end{array}$
\end{center}
\caption{(Color online) (a) Different components of the displacement-displacement correlation matrix $\mathbf{\Psi}(r)$ of an inherent structure containing primarily bound dislocations at $p=0.17$ with $N=10,800$ particles. (b)  Different components of the displacement-displacement correlation matrix $\mathbf{\Psi}(r)$ of an inherent structure containing primarily free dislocations at $p=0.04$ with $N=10,800$ particles. (c) Different components of the displacement-displacement correlation matrix $\mathbf{\Psi}(r)$ of an inherent structure containing disclinations at $q=0.015$ with $N=10,800$ particles.} \label{fig_8}
\end{figure*}

We then look at the cases where a substantial amount of topological defects are affecting the structures. Specifically, in Fig. \ref{fig_7} we show the spatial distribution of the vector displacement field $\mathbf{u}(\mathbf{r})$ for three representative examples: an inherent structure containing primarily bound dislocations at $p=0.17$, an inherent structure containing primarily free dislocations at $p=0.04$, and an inherent structure containing disclinations at $q=0.015$. All of the three fields in Fig. \ref{fig_7} appear to be approximately isotropic, suggesting a finite $\Psi(0)$. We also compute the different components of displacement-displacement correlation matrix $\Psi(r)$ for all these three cases, and the results are shown in Fig. \ref{fig_8}. The orthogonal components $\Psi_{12}$ appear to vanish, and the diagonal components are roughly the same, i.e., $\Psi_{11}(r) \approx \Psi_{22}(r)$, which satisfies the condition $\Psi_{\mu\nu}(\mathbf{r}) = \delta_{\mu\nu} \Psi(\mathbf{r})$ in our theory. In addition, the diagonal components $\Psi_{11}$ and $\Psi_{22}$ in Fig. \ref{fig_8} show relatively short-ranged correlations, which are consistent with the visualizations in Fig. \ref{fig_7}.

With the assumptions in our theory largely satisfied by our numerical examples in Sec. III, we proceed to investigate whether the numerically determined small-$k$ behavior of $S(k)$ matches the prediction by our theory. Indeed, as shown in Figs. \ref{fig_3}-\ref{fig_5}, at low and intermediate defect concentrations $p$ and $q$, the scaling exponent $\alpha$ in $S(k) \sim k^\alpha$ at small $k$ oscillates around 2, matching the quadratic scaling predicted by our theory; however, at very small or large $p$ and $q$ (relative to saturation), the exponent $\alpha$ slightly deviates from 2. The deviation at very small $p$ or $q$ can be attributed to the fact that at these defect concentrations the systems are not entirely homogeneous, which degrades the accuracy of our continuum-theory prediction; on the other hand, at large $p$ or $q$, the topological defects begin to interact with each other and modify the inherent structures accordingly, which is not taken into account in our continuum theory.

\section{Conclusions and Discussion}

In this work, we made an attempt to elucidate a possible mechanism for the observed hyperuniformity in disordered inherent structures of a wide spectrum of systems. In particular, we considered a representative class of disordered inherent structures which are linked to an original crystal state via continuous topological transformations involving dislocations and disclinations. We show via both numerical simulations and theoretical analysis that these topological transformations preserve the class-I hyperuniformity of the triangular lattice, and the structure factor $S(k)$ possesses a universal quadratic scaling as $k$ decreases at small $k$ at low defect concentrations.


Our continuum theory connects the large-scale density fluctuations in these inherent structures to the elastic displacement fields resulted from the topological defects. It indicates that class-I hyperuniformity can be preserved as long as the displacement fields resulted from individual defects decay fast enough from the source (i.e., the volume integrals of the displacements and squared displacements caused by individual defect are finite) and the displacement-displacement correlation matrix of the system is diagonalized and isotropic. Conceptually, the introduction of topological defects into a crystal does not affect the average particle density of the system (since the total number of particles are conserved), and any change in density fluctuations of the resulting disordered inherent structures could only come from the elastic displacement fields caused by the topological defects. As long as these elastic fields are homogenized and sufficiently localized, the salient features of large-scale density fluctuations of the original crystal, i.e., hyperuniformity, should be preserved. These results suggest promising new venues for the discovery, design, and generation of novel disordered hyperuniform materials.

Moreover, the inherent structures containing dislocations and disclinations studied in this work are quite different from the equilibrium structures containing the same type of topological defects in colloidal systems during 2D melting \cite{Za99, Wi11} in terms of various structural features, in particular the hyperuniformity and the large-$r$ scaling behavior of $g_2(r)$ and $C_6(r)$. These differences suggest that not only the types of defects, but also the spatial correlations of defects are key to understand the impact of defects on the structural features of crystalline systems. These results also highlights the cloaking effect of thermal fluctuation on hyperuniformity, and indicates that when looking for disordered hyperuniformity, in many cases one should probably look at potential-energy minima (local or global), which are only functions of the positional degrees of freedom and not affected by the vibrational degrees of freedom.

Here we studied the introduction of topological defects into triangular lattice, but given the duality of triangular lattice and honeycomb lattice, the dual of the various inherent structures obtained in this work are similar to the network structures obtained previously \cite{Ch21} that are topologically transformed from the honeycomb lattice through the introduction of Stone-Wales (SW) defects. Interestingly, those defected honeycomb network structures were found to capture the salient structural features of amorphous graphene and other 2D materials \cite{Ch21}. However, we note that previously it was demonstrated that as the SW defect concentration reaches certain critical value, the corresponding systems changes from class-I hyperuniformity to class-II hyperuniformity, which can be attributed to the fact that in those systems the bond angles were also regulated because of the underlying chemical constraints \cite{Ch21} and the additional coupling between defects likely modifies the behavior of large-scale density fluctuations and leads to the change in the class of hyperuniformity at sufficiently large defect concentrations. 

It is also noteworthy that topological defects such as dislocations and disclinations appear as point defects in two dimensions, while in three dimensions they are known to appear as line defects. Given this difference, it would be interesting to explore the impact of topological defects on the large-scale structural features of crystals in three dimensions, and the potential findings may shed light on the emergence of hyperuniformity in certain disordered inherent structures. For example, there were theories \cite{Ne85} suggesting that MRJ packings (or random close packings as termed by many experimentalists) might be considered as disordered topological variants of the tetrahedral particle packings, just like that Frank-Kasper phases \cite{Ne85, Re18, Ba20} are considered as ordered topological variants. 


\begin{acknowledgments}
We are grateful for Dr. Jaeuk Kim for very helpful discussion. 
\end{acknowledgments}


\end{document}